\documentclass[]{aa}  

\usepackage{natbib}
\bibpunct{(}{)}{;}{a}{}{,}
\usepackage{graphicx}
\usepackage{revsymb}	
\usepackage{amsmath}	
\usepackage[usenames]{color}	
\usepackage{txfonts}
\usepackage{amssymb}
\usepackage{color}
\usepackage{geometry} 
\usepackage[parfill]{parskip} 
\usepackage{amssymb}
\usepackage{slantsc}
\usepackage{array}
\usepackage{url}
\usepackage{amsfonts}
\usepackage[colorlinks=true,linkcolor=black, urlcolor=black, citecolor=black]{hyperref}
\usepackage{hyperref} 
\usepackage{sidecap}
\usepackage{graphicx}
\usepackage{xcolor}
\usepackage{nicefrac}
\usepackage{subfigure}
\usepackage{epsfig}
\colorlet{rouge}{red!70!darkgray}

\begin{document}
\title{Seismic Solar Models from Ledoux discriminant inversions}
\author{G. Buldgen\inst{1} \and P. Eggenberger\inst{1} \and V.A. Baturin\inst{2} \and T. Corbard\inst{3} \and J.Christensen-Dalsgaard\inst{4} \and S. J. A. J. Salmon\inst{1,5}\and A. Noels\inst{5} \and A. V. Oreshina\inst{2} \and R. Scuflaire\inst{5}}
\institute{Observatoire de Genève, Université de Genève, 51 Ch. Des Maillettes, CH$-$1290 Sauverny, Suisse \and Sternberg Astronomical Institute, Lomonosov Moscow State University, 119234 Moscow, Russia \and Université Côte d'Azur, Observatoire de la Côte d'Azur, CNRS, Laboratoire Lagrange, France \and Stellar Astrophysics Centre and Department of Physics and Astronomy, Aarhus University, DK-8000 Aarhus C, Denmark \and STAR Institute, Université de Liège, Allée du Six Août 19C, B$-$4000 Liège, Belgium}
\date{January, 2020}
\abstract{The Sun constitutes an excellent laboratory of fundamental physics. With the advent of helioseismology, we were able to probe its internal layers with unprecendented precision and thoroughness. However, the current state of solar modelling is still stained by tedious issues. One of these central problems is related to the disagreement between models computed with recent photospheric abundances and helioseismic constraints. The observed discrepancies raise questions on some fundamental ingredients entering the computation of solar and stellar evolution models.}
{We use solar evolutionary models as initial conditions for reintegrations of their structure using Ledoux discriminant inversions. The resulting models are defined as seismic solar models, satisfying the equations of hydrostatic equilibrium. These seismic models will allow us to better constrain the internal structure of the Sun and provide complementary information to that of calibrated standard and non-standard models.}{We use inversions of the Ledoux discriminant to reintegrate seismic solar models satisfying the equations of hydrostatic equilibrium. These seismic models were computed using various reference models with different equations of state, abundances and opacity tables. We check the robustness of our approach by confirming the good agreement of our seismic models in terms of sound speed, density and entropy proxy inversions as well as frequency-separation ratios of low-degree pressure modes.}{Our method allows us to determine with an excellent accuracy the Ledoux discriminant profile of the Sun and compute full profiles of this quantity. Our seismic models show an agreement with seismic data of $\approx 0.1\%$ in sound speed, density and entropy proxy after $7$ iterations in addition to an excellent agreement with the observed frequency-separation ratios. They surpass all standard and non-standard evolutionary models including ad-hoc modifications of their physical ingredients aiming at reproducing helioseismic constraints.}{The obtained seismic Ledoux discriminant profile as well as the full consistent structure obtained from our reconstruction procedure paves the way for renewed attempts at constraining the solar modelling problem and the missing physical processes acting in the solar interior by breaking free from the hypotheses of evolutionary models.}
\maketitle
\section{Introduction}

Over the course of the $20^{\rm{th}}$ century, the field of helioseismology has encountered major successes and has provided us highly precise measurements of the internal properties of the Sun. Thanks to the exquisite observational data taken over decades, seismology of the Sun has allowed to determine precisely the position of the base of the solar convective zone \citep{JCD91Conv, KosovBCZ, Basu97BCZ}, to measure the current helium abundance in the convective zone \citep{Vorontsov91,Dziembowski91, Antia94, BasuYSun, RichardY} as well as the $2$D profile of the rotational velocity \citep{BrownRota, Thompson1996, Howe2009} and the radial profile of structural quantities inside the Sun such as sound speed and density \citep[e.g.][for some illustrations including non-linear techniques and the first direct inversion of sound-speed from the asymptotic expression of pressure modes]{JCD1985,Antia94Nonlin, Marchenkov}. These results of unprecedented quality are at the origin of key questions for solar and stellar physics, showing the crucial role of the Sun and stars as laboratories of fundamental physics. 

Amongst them, the revision of the solar abundances starting almost two decades ago and culminating in \citet{AGSS09} caused a crisis in the solar modelling community that still awaits a definitive solution \citep[see e.g.][and references therein for additional discussions]{Antia05, Guzik06, Montalban06, Zaatri2007, SerenelliComp, Guzik2010, Bergemann2014, Zhang14}. The recent experimental measurement of opacity by \citet{Bailey} and \citet{Nagayama2019} confirmed the suspicions of the community that the source of the observed discrepancies with revised abundances could stem from the theoretical opacity tables used for solar models. However, while these certainly represent the main suspect, other contributors may well have a non-negligible impact in the total quantitative analysis of the mismatch between seismic data and evolutionary models of the Sun. 

Recently, \citet{Buldgen2019} analysed in depth the different contributors to the solar modelling problem using a combination of inversion techniques. They also showed that none of the current combinations of physical ingredients could   restore the agreement between low-metallicity standard evolutionary solar models and helioseismic constraints to the level of high-metallicity standard evolutionary solar models. Similarly to earlier studies \citep{Basu08,JCD2010, Ayukov2011, Ayukov2017, JCD18}, they concluded that a local increase in opacity was also insufficient to restore the agreement of low-metallicity evolutionary models with helioseismic data. They also analysed in depth the interplay between different physical ingredients such as the hypotheses made when computing microscopic diffusion, the equation of state and the formalism used for convection and showed that they could lead to small but significant differences at the level of precision expected from helioseismic inferences. 

Consequently, further constraining the required changes to solve the solar modelling problem might require to step out of the framework of evolutionary models, trying to provide direct seismic constraints on the possible inaccuracies of microphysical ingredients as well as macroscopic processes not included in the current standard solar models. To do so, a promising approach is to try to rebuild the solar structure as seen from helioseismic data, as done in e.g. \citet{Shibahashi1995}, \citet{Shibahashi1996}, \citet{Basu1996} , \citet{Gough2001}, \citet{Gough2004} and \citet{Turck2004}, and use this static structure to provide insights on potential revision of key ingredients of solar and stellar models. In this paper, we present a new approach to rebuild the solar structure based on inversions of the Ledoux discriminant, defined as 
\begin{align}
A=&\frac{1}{\Gamma_{1}}\frac{d \ln P}{d \ln r}-\frac{d \ln \rho}{d \ln r}.
\end{align} 

We demonstrate that the procedure converges on a unique solution for the Ledoux discriminant profile after only a few iterations and that the final structure for this new ``seismic Sun'' also agrees very well with all other structural inversions such as those of density, sound speed and entropy proxy defined in \citet{BuldgenS} as 
\begin{align}
S_{5/3}=\frac{P}{\rho^{5/3}}. 
\end{align}

A main advantage of the reconstruction procedure is that it provides a full profile of the Ledoux discriminant without the need of numerical differentiation. However, this does not mean that the method is devoid of interpolations when constructing the seismic models.

The goal of our reconstruction procedure is to provide a clearer insight as to the requirements of a revision of the opacity tables in solar conditions, which are the subject of discussion in the opacity community \citep[see e.g.][ and references therein]{Iglesias2015,Nahar2016,Blancard2016,Iglesias2017,Pain2018,Pradhan2018,Zhao2018,Pain2019,Pain2020} following the experimental measurements of \citet{Bailey} and \citet{Nagayama2019}. Consequently, we mainly focus on the reproduction of the profile of structural quantities in the radiative region of the Sun, where uncertainties in the solar $\Gamma_{1}$ profile may be considered negligible in the $A$ profile determined from the inversion.

In addition, it can be shown that the behaviour of the $A$ profile in the deep radiative layers is mostly determined by the temperature gradient, the mean molecular weight gradient only contributing to the expression of $A$ near the base of the convective zone and in regions affected by nuclear reactions. Consequently, we can have a direct measurement of the temperature gradient in these regions and directly quantify the required changes of opacity for various chemical compositions and underlying equations of state, once the internal structure has been reliably determined. Such a determination is complementary to the approaches used for example in \citet{JCD2010}, \citet{Ayukov2017} and \citet{Buldgen2019}, where ad-hoc modifications were applied in calibrated solar models. These measurements are to be compared to the expected revisions of theoretical opacity computations and the current experimental measurements available, helping guide a revision of standard ingredients of solar models by providing additional ``experimental'' opacity estimations directly from helioseismic observations \footnote{We note that similar data are given in \citet{Gough2004}, page 14, but was unfortunately provided before the revision of the abundances and the subsequent appearance of the so-called ``solar modelling problem''.}.

Near the base of the solar convective zone, our approach will provide a complementary method to that of \citet{JCD18} in studying the hydrostatic structure of the solar tachocline\footnote{This is the region marking the transition between the latitudinal differential rotation in the solar convection enevelope and the rigid rotation of the radiative interior and may be subject to extra mixing.} \citep{Spiegel1992}. Indeed, the properties of the mean molecular weight gradients in this region play a key role in understanding the angular momentum transport mechanisms acting in the solar interior \citep{Gough1998,Spruit1999,Charbonnel2005,Eggenberger2005,Spada2010,Eggenberger2019} that also now cause troubles in understanding the rotational properties of main-sequence and evolved low-mass stars observed by \textit{Kepler} \citep{Deheuvels2012, Mosser2012, Deheuvels2014, Lund2014, Benomar2015, Nielsen2017}. However, as we mention below, the finite resolution of the inversion techique leads to higher uncertainties that may require further adaptations.

We start in Sect. \ref{SecMethod} by presenting the reconstruction procedures, including the choice of reference models. In Sect. \ref{SecVerif}, we discuss the agreement with other structural inversions and the origins of the remaining discrepancies. The limitations and further dependencies on initial conditions are discussed in Sects. \ref{SecDisc} and \ref{SecLimitations}, while perspectives for future applications are presented in Sect. \ref{SecConc}.

\section{Methodology}\label{SecMethod}

In this Section, we present our approach to construct a static structure of the Sun in agreement with seismic inversions of the Ledoux discriminant. We start by presenting the sample of reference evolutionary models we use for the reconstruction procedure as well as their various physical properties. Using different reference models allows us to determine the amount of "model dependency" remaining in the final computed structure, which has to be taken into account in the total uncertainty budget when discussing solar properties. We then present the inversion procedure and how it can be used iteratively to reconstruct a full ``seismic model'' of the Sun. 

\subsection{Sample of reference models}

The use of the linear variational relations that form the basis to carry out structural inversions of the Sun requires that a suitable reference model be computed beforehand. This is ensured in our study by following the usual approach to calibrate solar models. In other words, we use stellar models of $1~\rm{M}_{\odot}$, evolved to the solar age and reproducing at this age the solar radius and luminosity, taken here from \citet{Prsa2016}.

All our models include the transport of chemical elements by microscopic diffusion and are constrained to reproduce a given value of $\left(\mathit{Z}/\mathit{X}\right)_{\odot}$ at the solar age. However, they are not ``standard'' in the usual sense of the word, as the $\left(\mathit{Z}/\mathit{X}\right)_{\odot}$ used as a constraint in the calibration is not necessarily consistent with the reference abundance tables. In this study, we use the GN93 and AGSS09 chemical abundances \citep{GrevNoels, AGSS09} including for some models the recent revision of neon abundance determined by \citet{Landi} and \citet{Young}, denoted AGSS09Ne. The $\left(\mathit{Z}/\mathit{X}\right)_{\odot}$ value used for the calibrations spans the range allowed by the AGSS09 and the GN93 tables. In other words, some models reproduce the ${\left(\mathit{Z}/\mathit{X}\right)}_{\odot}$ value from the GN93 abundance tables while including the AGSS09 abundance ratios of the individual elements. This allows to test a wider ranges of initial structures for our procedure while still remaining within the applicability range of the linear variational equations. 

We considered variations of the following ingredients in the calibration procedure: equation of state, formalism of convection, opacity tables and $\mathit{T}(\tau)$ relations for the atmosphere models. Namely, we considered the FreeEOS \citep{Irwin} and the SAHA-S equations of state \citep{Gryaznov04, Gryaznov06, Gryaznov13, Baturin}, the OPAL \citep{OPAL}, OPLIB \citep{Colgan} and OPAS \citep{Mondet} opacity tables, the MLT \citep{Cox} and FST \citep{Canuto91, Canuto, Canuto96} formalisms for convection and the Vernazza \citep{Vernazza}, Krishna-Swamy \citep{Krishna1966} and Eddington $\mathit{T}(\tau)$ relations, denoted ``VAL-C'', ``K-S'' and ``Edd'' in Table \ref{tabSTDModels}. We considered the nuclear reaction rates of \citet{Adelberger}, the low temperature opacities of \citet{Ferguson} and the formalism of diffusion of \citet{Thoul} using the diffusion coefficients of \citet{Paquette}, taking into account the effects of partial ionization. 

The models have been computed with the Liège Stellar Evolution Code \citep[CLES,][]{ScuflaireCles} and their global properties are summarized in Table \ref{tabSTDModels}. A key parameter to the reconstruction procedure is the position of the base of the convective zone (BCZ), because, as shown in Sect. \ref{SecInversion}, it is not altered during the iterations. All other parameters are informative of the properties of the calibrated model but do not enter the reconstruction procedure. We can see that there is a clear connexion between the $\mathrm{m}_{0.75}$ and the position of the BCZ. The dichotomy in two families of models depending on their metallicity is also seen in the values of the $\mathrm{m}_{0.75}$ parameter. Indeed, low-$\mathit{Z}$ models (namely Models $4$, $5$ and $6$) will show a higher value of $\mathrm{m}_{0.75}$ associated also with a low density in the envelope, while high-$\mathit{Z}$ models (Models $1$ to $3$ and $7$ to $10$) will show a much denser enveloper and thus a lower value of $\mathrm{m}_{0.75}$ in better agreement with the solar value determined by \citet{Vorontsov13}. We denoted the reference models as Model $i$ while the final model will be denoted Sismo $i$, such that the Sismo 10 denotes the reconstructed model from the starting point denoted Model 10.

\begin{table*}[t]
\caption{Parameters of the reference models for the reconstruction}
\label{tabSTDModels}
  \centering
    \resizebox{\linewidth}{!}{%
\begin{tabular}{r | c | c | c | c | c | c | c | c | c | c }
\hline \hline
\textbf{Name}&\textbf{$\left(r/R\right)_{\rm{BCZ}}$}&\textbf{$\left( m/M \right)_{\rm{CZ}}$}&\textbf{$\mathit{Z}_{\rm{CZ}}$}&\textbf{$\mathit{Y}_{\rm{CZ}}$}&\textbf{$\mathrm{m}_{0.75}$} &\textbf{EOS}&\textbf{Opacity}&\textbf{Relative Abundances} & \textbf{Convection} & \textbf{Atmosphere}\\ \hline
Model 1&$0.7145$&$0.9762$& $0.01797$ & $0.2455$ & $0.9826$ & FreeEOS & OPAL & GN93 & MLT & VAL-C\\
Model 2&$0.7117$&$0.9751$& $0.01811$& $0.2394$ & $0.9822$ & FreeEOS & OPLIB & GN93 & MLT & VAL-C\\ 
Model 3&$0.7127$&$0.9751$& $0.01766$ & $0.2587$ & $0.9820$ & FreeEOS & OPAL & AGSS09Ne & MLT & VAL-C\\
Model 4&$0.7224$&$0.9785$& $\mathit{0.01389}$ & $\mathit{0.2395}$ & $0.9832$ & SAHA-S & OPAL & AGSS09Ne & MLT & VAL-C\\
Model 5&$0.7209$&$0.9784$& $\mathit{0.01395}$ & $\mathit{0.2358}$ & $0.9834$ & SAHA-S & OPAS & AGSS09 & MLT & VAL-C\\
Model 6&$0.7220$&$0.9788$& $\mathit{0.01362}$ & $\mathit{0.2337}$ & $0.9836$ & SAHA-S & OPAS & AGSS09 & MLT & VAL-C\\
Model 7&$0.7144$&$0.9756$& $0.01765$ & $0.2591$ & $0.9822$ & FreeEOS & OPAL & AGSS09 & MLT & VAL-C\\
Model 8&$0.7144$&$0.9762$& $0.01797$ & $0.2455$ & $0.9826$ & FreeEOS & OPAL & GN93 & MLT & KS\\
Model 9&$0.7144$&$0.9762$& $0.01797$ & $0.2454$ & $0.9826$ & FreeEOS & OPAL & GN93 & FST & KS\\
Model 10&$0.7145$&$0.9762$& $0.01797$ & $0.2455$ & $0.9826$ & FreeEOS & OPAL & GN93 & MLT & EDD\\
\hline
\end{tabular}
}
\small{\textit{Note:} Model 4, 5 and 6, with their composition in \textit{italics}, have been calibrated using $\left(Z/X\right)_{\odot}=0.0186$, while all other models use $\left(Z/X\right)_{\odot}=0.0244$. We use the following definitions: $\left(r/R\right)_{\rm{BCZ}}$ is the radial position of the BCZ in solar radii, $\left( m/M \right)_{\rm{CZ}}$ is the mass coordinate at the BCZ in solar masses, $\mathrm{m}_{0.75}$ is the mass coordinate at $0.75\mathrm{R_{\odot}}$ in solar masses, $\mathit{Y}_{\rm{CZ}}$ and $\mathit{Z}_{\rm{CZ}}$ are the helium and average heavy element mass fraction in the CZ. }
\end{table*}

\subsection{The reconstruction procedure}\label{SecInversion}

The starting point of the reconstruction procedure is a calibrated solar model, for which the linear variational relations can be applied. This implies that, following \citet{Dziemboswki90}, the relative frequency differences between the observed solar frequencies and those of the theoretical model can be related to corrections of structural variables as follows: 
\begin{align}
\frac{\delta \nu_{n,\ell}}{\nu_{n,\ell}}=\int_{0}^{R}K^{n,\ell}_{s_{1},s_{2}}\frac{\delta s_{1}}{s_{1}}dr + \int_{0}^{R}K^{n,\ell}_{s_{2},s_{1}}\frac{\delta s_{2}}{s_{2}}dr + \mathcal{F}_{\mathrm{Surf}}, \label{EqInversion}
\end{align}
with $\delta$ denoting here the relative differences between given quantities following
\begin{align}
\frac{\delta x}{x}=\frac{x_{\rm{Obs}}-x_{\rm{Ref}}}{x_{\rm{Ref}}},
\end{align}
where $x$ can be in our case a frequency, $\nu_{n,\ell}$ or the local value of a structural variables taken at a fixed radius such as e.g. $A$, $\rho$, $c^{2}=\frac{\Gamma_{1}P}{\rho}$ or $\Gamma_{1}=\left[\frac{\partial \ln P}{\partial \ln \rho}\right]_{S}$, denoted $s_{i}$. The subscripts ``$\rm{Ref}$'' and ``$\rm{Obs}$'' denote the theoretical values of the reference model and the observed solar values, respectively. 

In Eq.~(\ref{EqInversion}), the $K^{n,\ell}_{s_{i},s_{j}}$ are the so-called structural kernel functions which serve as ``basis functions'' to evaluate the structural corrections to a given model in an inversion procedure. The $\mathcal{F}_{\mathrm{Surf}}$ function denotes the surface correction term, which we model as a sum of inertia-weighted Legendre polynomials in frequency (up to the degree $6)$, with the weights determined during the inversion procedure, considering a dependency on frequency alone.

From \citet{Gough1993}, \citet{Kosovichev1993}, \citet{Elliott1996}, \citet{Kosovichev} and \citet{BuldgenKer}, we know that Eq.~(\ref{EqInversion}) can be written for a wide range of variables appearing in the adiabatic pulsation equations. In what follows, we will focus on using the $\left(A,\Gamma_{1} \right)$ structural pair. 

In this study, the adiabatic oscillation frequencies have been computed using the Liège adiabatic Oscillation Code \citep[LOSC,][]{ScuflaireOsc}, the structural kernels and the inversions have been computed using an adapted version of the InversionKit software and the Substractive Optimally Localized Average (SOLA) inversion technique \citep{Pijpers}. The frequency dataset considered is a combination of MDI and BiSON data from \citet{BasuSun} and \citet{Davies}. The trade-off parameters of the inversions were adjusted following the guidelines of \citet{RabelloParam}.

The reconstruction procedure of a seismic solar model is done as follows (more details are given in Appendix \ref{SecNumDetails}):
\begin{enumerate}
\item The corrections to the Ledoux discriminant for the reference model, $\delta A$ are determined using the SOLA method. 
\item The $A$ profile of the model is corrected such that $A^{'}=A+\delta A$ in the radiative mantle of the model, namely between $0.08\rm{R}_{\rm{\odot}}$ and the BCZ of the model. 
\item The structure of the model is then reintegrated, assuming hydrostatic equilibrium and assuming no changes in mass and radius, using the corrected Ledoux discriminant $A^{'}$ and leaving $\Gamma_{1}$ untouched. 
\item The corrected model then becomes the reference model for a new inversion in Step $1$.
\end{enumerate}
The procedure is stopped once no significant corrections can be made to the $A$ profile. This is typically reached after a few iterations ($\approx 7$). This limit is determined by the dataset used for the structural inversion as well as the inversion technique itself. Indeed, the dataset will determine the capabilities of the inversion technique to detect mismatches between the reference model and the solar structure from a physical point of view, while the inversion technique itself will be limited by its intrinsic numerical capabilities. 

For example, a limitation of the dataset is the impossibility of p-modes to probe the deepest region of the solar core. In our case, we considered, following a conservative approach, that the Ledoux discriminant inversion did not provide reliable information below $0.08~\rm{R}_{\rm{\odot}}$, because of the poor localisation of the averaging kernels in that region. Another example of limitation of the SOLA method is illustrated in the tachocline region, and could already be seen in Fig. $28$ from \citet{Kosovichev2011} and Fig. $2$ from \citet{BuldgenA}. Due to the approach chosen to solve the integral equations by determining a localized average, the SOLA method is not well-suited for determining corrections in regions of sharp transitions \citep[see e.g.][for a discussion on the finite resolution of inversions near the base of the convective zone]{JCD1985,JCD1989, JCD91Conv}. Similarly, the Regularized Least Square (RLS) technique using the classical Tikhonov regularisation \citep{Tikhonov1963} will suffer from similar limitations. One potential solution to the issue would be to use a non-linear RLS technique, allowing sharp variations of the inversion results, as done in \citet{corbard99} for the solar rotation profile in the tachocline. 

The procedure is thus quite straightforward from a numerical point of view, but there are a few details that require some additional discussion. The fact that we stop correcting the Ledoux discriminant profile around $0.08~\rm{R}_{\rm{\odot}}$ can lead to spurious behaviour if no proper reconnection with the reference profile is performed. To avoid this, we carry out a cubic interpolation between the corrected $A^{'}$ and the $A$ on a small number of points, starting at the point of lowest correction in $A$ around $0.08~\rm{R}_{\rm{\odot}}$.

\begin{figure*}
	\centering
		\includegraphics[width=17cm]{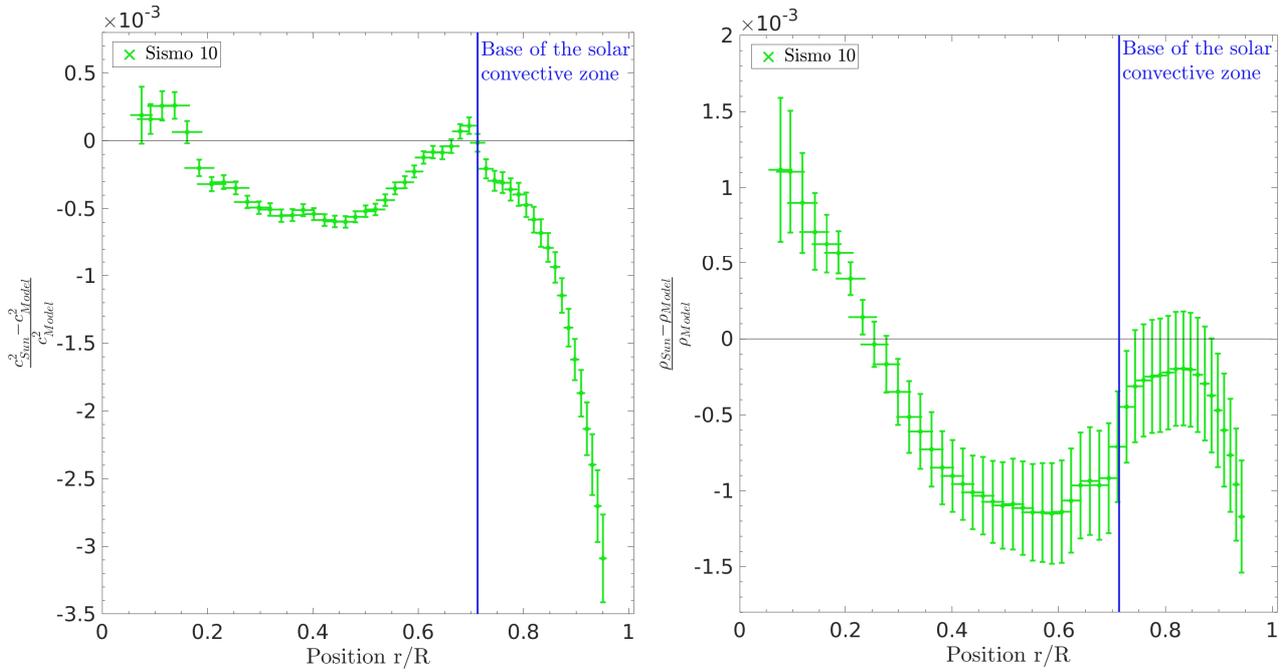}
	\caption{Left panel: Relative differences in squared adiabatic sound speed between the Sun and the Seismic model $10$ (Sismo 10). Right panel: Relative differences in density between the Sun and the Seismic model $10$ (Sismo 10).}
		\label{Figc2SunMod1}
\end{figure*}

\begin{figure*}
	\centering
		\includegraphics[width=17cm]{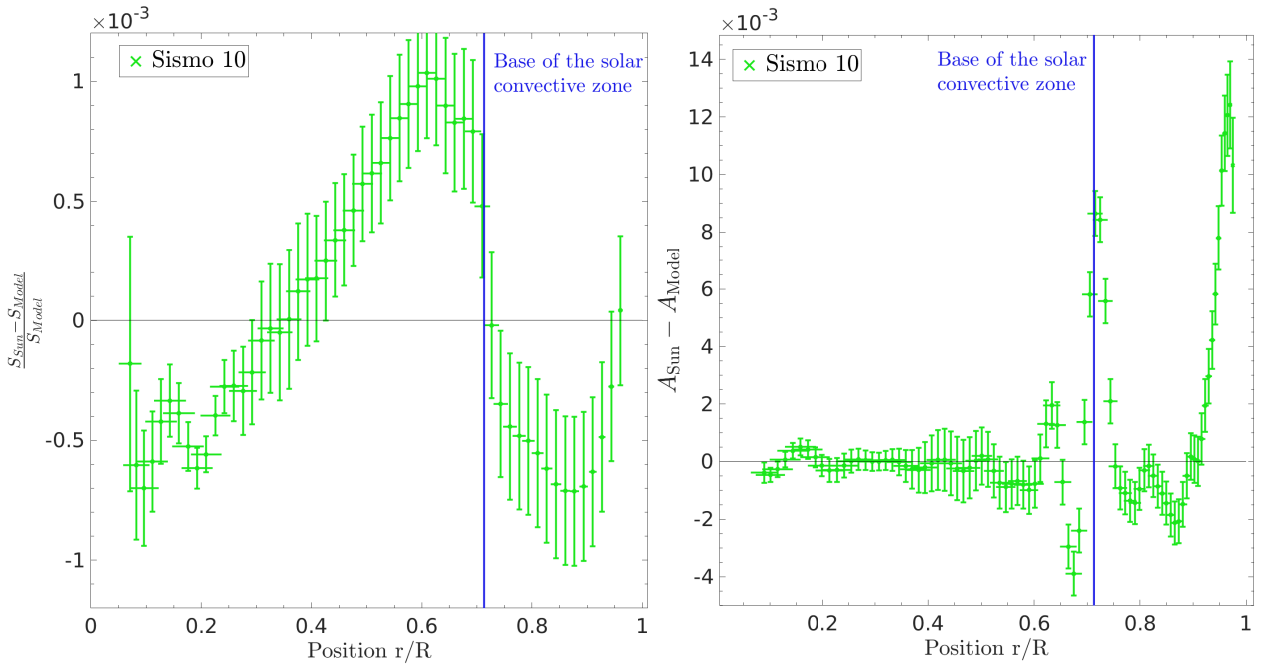}
	\caption{Left panel: Relatives differences in entropy proxy, $S_{5/3}$ between the Sun and the Seismic model $10$ (Sismo 10). Right panel: Differences in Ledoux discriminant between the Sun and the Seismic model $10$ (Sismo 10).}
		\label{FigASunMod1}
\end{figure*}

In the convective zone, no correction to the $A$ profile is applied as the inversion results are not trustworthy. Indeed, the inversion has a tendency to overestimate the amplitude of the corrections in a region where $A$ is very small, as a consequence of the low amplitude of both chemical gradients and departures from the adiabatic temperature gradient in the lower parts of the convective zone. Consequently, the corrections in the convective zone are actually implicitly applied when the structure is reintegrated to satisfy hydrostatic equilibrium with the boundary conditions on M and R. Thus, despite not directly correcting the structural variables in the convective zone with the inversion, we are still able to significantly improve the agreement of sound speed, density, Ledoux discriminant and entropy proxy inversions for the reconstructed models, as illustrated in the left and right panels of Figs. \ref{Figc2SunMod1} and \ref{FigASunMod1} for Model 10 after $7$ iterations.

As we mentioned above, the reconstruction procedure does not explicitly apply corrections in $A$ in the convective layers. This means that the corrections in these regions are a sole consequence of the modifications required to satisfy mass conservation in the reconstructed models. As we will see in Sect. \ref{SecSoundSpeed} when looking at the changes in squared adiabatic sound speed at each iteration, the agreement in the convective envelope for this specific quantity is actually not improved over the reconstruction procedure. Consequently, it is clear that our seismic models do not provide an as good agreement in adiabatic sound speed in those regions as those determined in previous studies explicitly correcting the profiles in the convective envelope \citep[see e.g.][]{Antia94Nonlin, Turck2004, VorontsovSolarEnv2014}. This is a direct consequence of our reconstruction method, for which we chose to focus on the deeper radiative layers, where the uncertainties on $\Gamma_{1}$ are much smaller. Thus, in comparison to previous studies, our models perform very well in the deeper layers, especially for the density profile. This improvement of the density and entropy proxy profiles over the course of the iterations is a direct consequence of the mass conservation. Indeed, even if the convective layers are left untouched, the variations of density resulting from the $A$ corrections applied in the radiative interior will be compensated by larger variations in the upper, less-dense, convective layers, leading to an improvement of the agreement with the Sun for both density and entropy proxy.

At the exact location of the BCZ, the corrections to the Ledoux discriminant are still applied, using the $\delta A$ amplitude given by the inversion point with the closest central value of the averaging kernels. In the cases considered here, this only implies a minimal shift, well within both the vertical and resolution error bars of the inversion.

We thus define at this location a unique set of thermodynamical variables, $\rho$, $P$ and $\Gamma_{1}$ that will define the properties of the lower convective envelope in hydrostatic equilibrium. However, since the radial position of the transition between convective and radiative regions determined from the Schwarzschild criterion is not modified in the reconstruction, the determined seismic model will not follow exactly the same density profiles in the convective envelope, as we will see later. In the deep convective layers, the behaviour of the seismic model will be essentially determined by the original $A$ and $\Gamma_{1}$ profiles as well as the satisfaction of hydrostatic equilibrium through the determination of $m(r)$ at some radius.

Another region left untouched in the reconstruction procedure are the surface layers of the model, namely the substantially super-adiabatic convective layers as well as the atmosphere. This choice is justified by the fact that inversions based on the variational principle of adiabatic stellar oscillations are unable to provide reliable constraints in these regions; thus the inferred corrections would not be appropriate. This also justifies the use of different atmosphere models and formalisms of convection, as we can then directly measure their impact on the final reconstructed structure of the Sun (see Sect. \ref{SecDisc}).    

\begin{figure*}
	\centering
		\includegraphics[width=16cm]{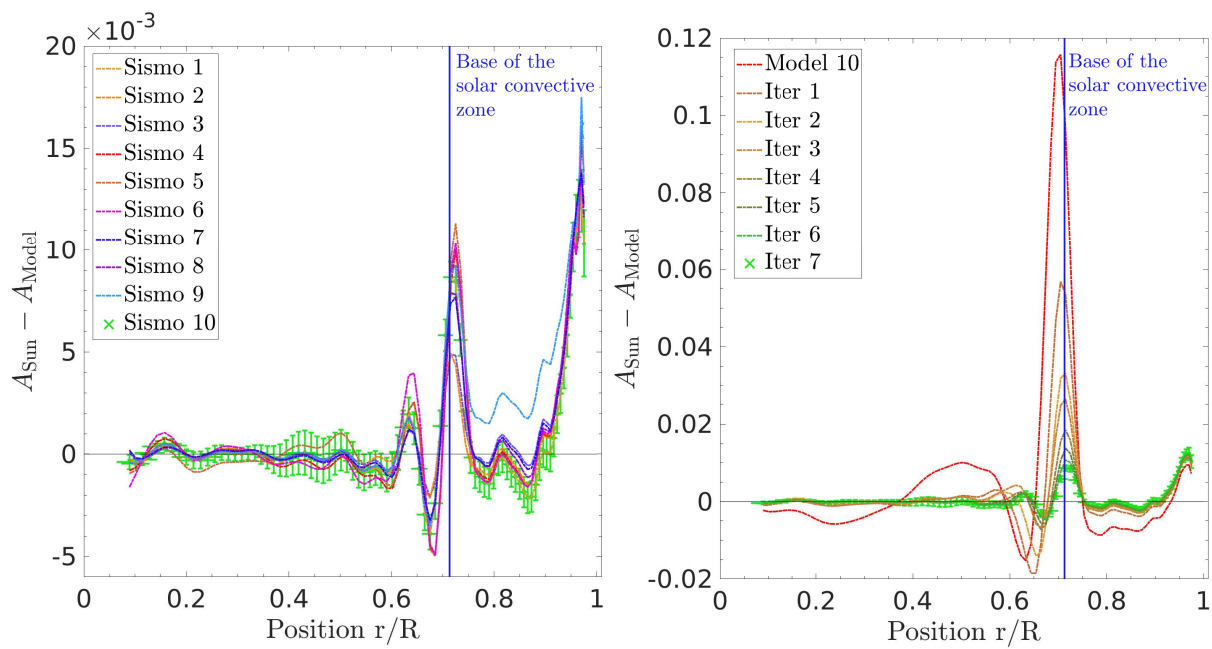}
	\caption{Left panel: Illustration of the agreement in Ledoux discriminant for the reconstructed models using the reference models of Table \ref{tabSTDModels} as initial conditions. Right panel: Illustration of the convergence of the $A$ corrections for successive iterations of the reconstruction procedure in the case of Model 10.}
		\label{FigAAll}
\end{figure*}

In the right panel Fig. \ref{FigAAll}, we illustrate the convergence of the reconstruction procedure for Model $10$. The final agreement in Ledoux discriminant inversions for all models in our sample is illustrated in the left panel of Fig. \ref{FigAAll} and a selection of the corresponding Ledoux discriminant profiles are illustrated in Fig. \ref{FigASunRec}. As can be seen, the agreement is excellent in the deep radiative layer, whatever the initial conditions. Small discrepancies can be seen just below the BCZ at the resolution limit of the SOLA inversion. The central regions (below $0.08~\mathrm{R}_{\odot}$) are also slightly different, since they are not modified during the reconstruction procedure. 

From the analysis of the Ledoux discriminant inversions, we can conclude that the reconstruction procedure has been able to provide a precise, model-independent profile of this quantity in the Sun which can be used to analyse the limitations of the current models. As mentioned previously, another key aspects for future advances in helioseismology is the potential observation of gravity modes. Below $\approx \rm{200~\mu Hz}$, the gravity modes follow a regular pattern and are described to the first order by an asymptotic expression as a function of their period, $P_{n,\ell}$

\begin{align}
P_{n,\ell}=\frac{P_{0}}{\sqrt{\ell (\ell +1)}} \left( n +\ell/2 +\upsilon \right), \label{EqGmodes}
\end{align}
with $n$ the radial order, $\ell$ the degree, $\upsilon$ a phase shift depending on the properties near the BCZ and $P_{0}$ is defined by
\begin{align}
P_{0}=\frac{2 \pi^{2}}{\int_{0}^{r_{\rm{BCZ}}} \left( N/r\right) dr}=\frac{2 \pi^{2}}{\int_{0}^{r_{\rm{BCZ}}} \sqrt{\vert gA/r^{2}\vert} dr},
\end{align}
with $N$ the Brunt-Väisälä frequency and $r_{\rm{BCZ}}$ the radial position of the BCZ. This implies that the separation between two modes of consecutive $n$ in the asymptotic regime, denoted the asymptotic period spacing, is determined by $P_{0}$ (i.e. by the integral of the Brunt-Väisälä frequency up to the BCZ).

We illustrate the results of these integrals of $P_{0}$  for our seismic and reference models in Table \ref{tabSpacing}, which shows that despite having good constraints on the Ledoux discriminant in the deep radiative layer of the Sun, the fact that we are missing the deep core still allows for a significant variation of the period spacing of g-modes in our sample of reconstructed structures. Typically, we find a range of period spacing values spanning an interval of $25~s$ far from the observed value of $2040~s$ of \citet{Fossat} but closer to the theoretical one of $2105~s$ of \citet{Provost2000}. Only slight changes in the period spacing values, of the order of $10~s$, are found if the reconstruction procedure is carried out using GOLF data from \citet{Salabert2015} instead of BiSON data for the low degree modes. However, these determined values can be significantly changed by altering the $A$ profile below $0.08~\mathrm{R}_{\odot}$ without destroying the agreement with the inversion results from p-modes. Hence, another input is required to better constrain the expected period spacing of the solar gravity modes. This will be further discussed in Sect. \ref{SecDisc} when studying the impact of the choice of the reference model. 

\begin{table}[t]
\caption{Comparison between period spacing values for the seismic models after 7 iterations and the reference models values.}
\label{tabSpacing}
  \centering
\begin{tabular}{r | c | c}
\hline \hline
\textbf{Name}&\textbf{$P_{0}$ ($s$)}&\textbf{$P_{0,\rm{Ref}}$ ($s$)} \\ \hline
Sismo 1&$2162$&$2162$\\
Sismo 2&$2161$&$2138$\\
Sismo 3&$2157$&$2157$\\
Sismo 4&$2170$&$2186$\\
Sismo 5&$2175$&$2207$\\
Sismo 6&$2176$&$2199$\\ 
Sismo 7&$2155$&$2160$\\
Sismo 8&$2164$&$2163$\\
Sismo 9&$2153$&$2188$\\
Sismo 10&$2164$&$2164$\\
\hline
\end{tabular}

\end{table}

\section{Agreement with other seismic indices}\label{SecVerif}

In the previous section, we focused on demonstrating that using successive inversions of the Ledoux discriminant allowed us to determine a model-independent profile for this quantity. However, the main benefit of the reconstruction procedure is that it allows to determine fully consistent seismic models of the solar structure, that are also in excellent agreement in terms of other structural variables. In this section, we will show that once the reconstruction procedure has converged, we reach a level of agreement in the radiative zone of $\approx 0.1\%$ for structural inversions of $\rho$, $c^{2}$ and $S_{5/3}=P/\rho^{5/3}$. This implies that most of the issues with our depiction of the solar structure are clearly related to the radiative layers, as we are able to suppress efficiently other traces of mismatches by correcting $A$ in the radiative region. 

\subsection{Sound-speed inversions}\label{SecSoundSpeed}

\begin{figure*}
	\centering
		\includegraphics[width=16cm]{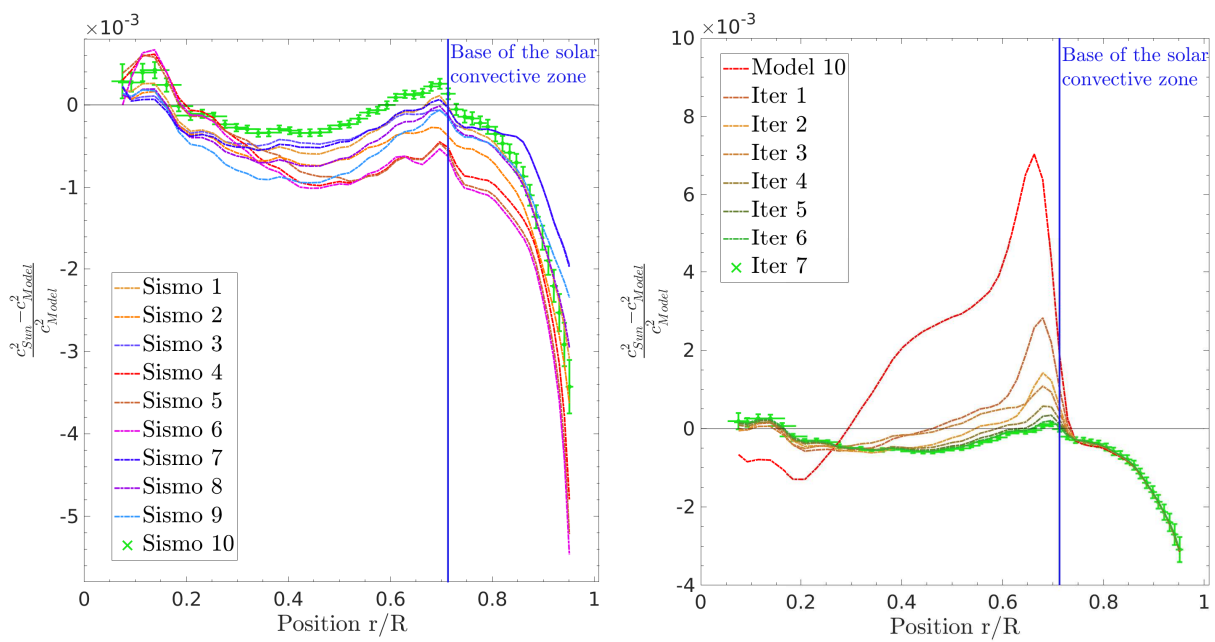}
	\caption{Left panel: Agreement in relative sound-speed differences for the reconstructed models using the reference models of Table \ref{tabSTDModels} as initial conditions. Right panel: Illustration of the convergence of the sound speed relative differences for successive iterations of the reconstruction procedure in the case of Model 10.}
		\label{Figc2All}
\end{figure*}

The sound-speed inversions are carried out using the $(c^{2},\rho)$ kernels in Eq.~(\ref{EqInversion}). As can be seen from the left panel of Fig. \ref{Figc2All}, the agreement for all models is of $\approx 0.1\%$ in the radiative region and the lower parts of the convective region. From the right panel of \ref{Figc2All}, we also see that, as mentioned earlier, the sound speed profile in the convective envelope is not corrected by the reconstruction procedure, as no corrections in $A$ are applied in the convective envelope.

These results confirm that, as expected, the solar modelling problem is mostly an issue related to the temperature gradient of the low-metallicity models in the radiative regions. However, from the analysis of the successive changes due to the iterations in the reconstruction procedure, illustrated in the right panel of Fig. \ref{Figc2All}, we can see that the bulk of the profile is corrected after the third reconstruction. The first step mostly corrects the sound-speed discrepancy in the radiative zone, near the BCZ. The remaining discrepancies are efficiently corrected in the following iterations. However, one can see that there is a remaining discrepancy even for the model obtained after $7$ iteration. These variations remain very small and are not linked to significant deviations in the $A$ profile. They do not seem to be linked to $\Gamma_{1}$ differences between the models, but rather to the position of the BCZ and the mass coordinate at that position. Indeed, these parameters determine the density profile of the models and it is clear that the models with the worst agreement on the position of the BCZ with respect to the helioseismic value of $0.713\pm 0.001~\mathrm{R_{\odot}}$ also have the largest discrepancies in both sound-speed and, as shown in Sect. \ref{SecDensity}, density profiles. This implies that an additional selection of the models based on their position of the BCZ and the mass coordinate at the BCZ could be used as a second step. Unfortunately, we do not have a direct measurement of the mass coordinate at the BCZ. \citet{VorontsovSolarEnv2014} were able to determine the mass coordinate at $0.75\mathrm{R_{\odot}}$, $\mathrm{m}_{0.75}=0.9822\pm0.0002~\mathrm{M_{\odot}}$ and discussed its importance as a calibrator of the specific entropy in the solar convective envelope. All our reconstructed models are in good agreement with the determined value of the $\mathrm{m}_{0.75}$ parameter of \citet{VorontsovSolarEnv2014}, as illustrated in Table \ref{tabM075}. Similarly, they all show very good agreement in entropy proxy inversions as we will further discuss in Sec. \ref{SecEntropy}.

\begin{table}[t]
\caption{Comparison between $\mathrm{m}_{0.75}$ for the seismic models after 7 iterations and their corresponding reference models values}
\label{tabM075}
  \centering
\begin{tabular}{r | c | c}
\hline \hline
\textbf{Name}&\textbf{$\mathrm{m}_{0.75}$}&\textbf{$\mathrm{m}_{0.75,\rm{Ref}}$} \\ \hline
Sismo 1&$0.9823$&$0.9826$\\
Sismo 2&$0.9823$&$0.9822$\\
Sismo 3&$0.9824$&$0.9820$\\
Sismo 4&$0.9823$&$0.9832$\\
Sismo 5&$0.9823$&$0.9834$\\
Sismo 6&$0.9823$&$0.9836$\\ 
Sismo 7&$0.9824$&$0.9822$\\
Sismo 8&$0.9824$&$0.9826$\\
Sismo 9&$0.9823$&$0.9826$\\
Sismo 10&$0.9823$&$0.9826$\\
\hline
\end{tabular}

\end{table}

\subsection{Entropy proxy inversions}\label{SecEntropy}

\begin{figure*}
	\centering
		\includegraphics[width=16cm]{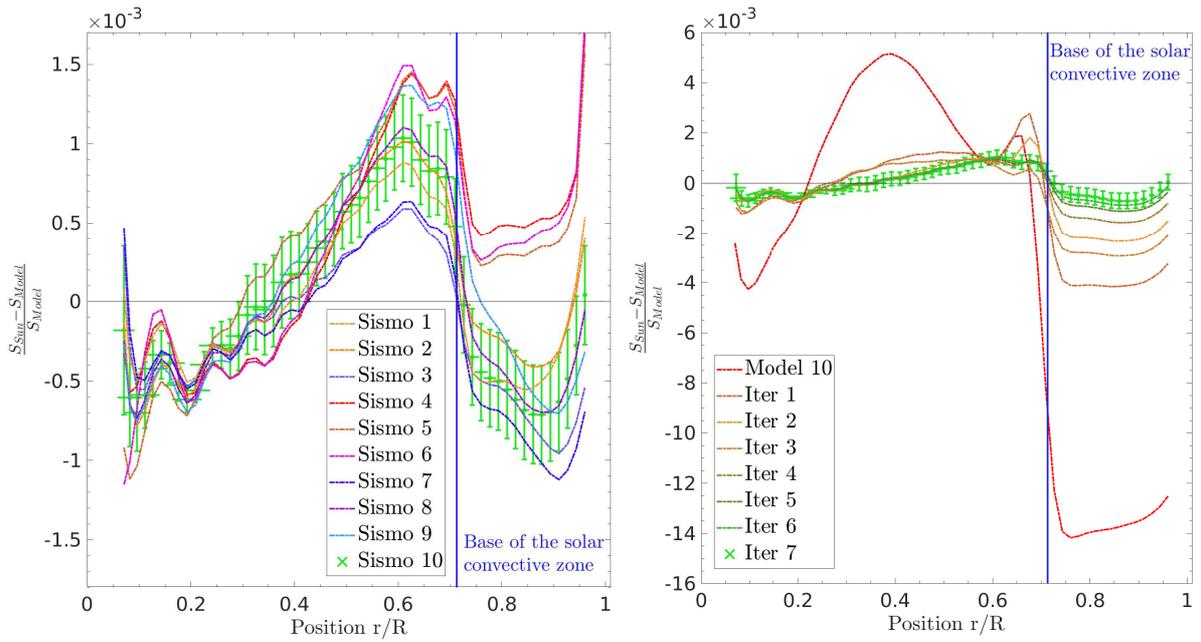}
	\caption{Left panel: Agreement in relative entropy proxy differences for the reconstructed models using the reference models of Table \ref{tabSTDModels} as initial conditions. Right panel: Illustration of the convergence of the entropy proxy relative differences for successive iterations of the reconstruction procedure in the case of Model 10.}
		\label{FigSAll}
\end{figure*}

The entropy proxy inversions are carried out using the $(S_{5/3},\Gamma_{1})$ kernels in Eq.~(\ref{EqInversion}). These kernels have been presented in \citet{BuldgenKer} and applied to the solar case in \citet{BuldgenS} and \citet{Buldgen2019}. From the left panel of Fig. \ref{FigSAll}, we can see that the agreement for this inversion is also of $\approx 0.1\%$ in the radiative and convective regions for all models, whatever their initial conditions. This confirms that correcting for the $A$ profile in the radiative region leads to an excellent agreement of the height of the plateau in the convective zone. 

From a closer analysis of the reconstruction procedure, we can see that a good agreement of the height of the plateau is reached after $3$ iterations. The explanation for this is found in the form of the applied corrections to the model. At first, the reconstruction procedure is dominated by the large discrepancies near the BCZ in all models. Once these differences have been partially corrected, we can see in the right panel of Fig. \ref{FigSAll} that the corrections in the deeper layer of the radiative zone remain significant for the second iteration. It then takes $2$ iterations to completely erase discrepancies in the $A$ profile below $0.6~\mathrm{R_{\odot}}$. Once this is achieved, the inferred $A$ profile shows an oscillatory behaviour in these regions (as seen in Fig. \ref{FigAAll}). This phenomenon is linked to a form of Gibbs phenomenon, due to the fact that the remaining deviations are located in very narrow region, just below the BCZ and are too sharp to be properly sampled by the classical SOLA method we use in this study. From a physical point of view, the remaining discrepancies may originate from multiple contributions: a slight mismatch between the transition from the radiative to convective outward transport of energy, which causes a sharp variation in $A$ and a mismatch in $A$ in the last percent of solar radii just below the BCZ. Of course, it should be recalled that a breaking of spherical symmetry in the structure of the tachocline region could also lead to mismatches  with any modelling assuming spherical symmetry. 

\subsection{Density inversions}\label{SecDensity}

\begin{figure*}
	\centering
		\includegraphics[width=16cm]{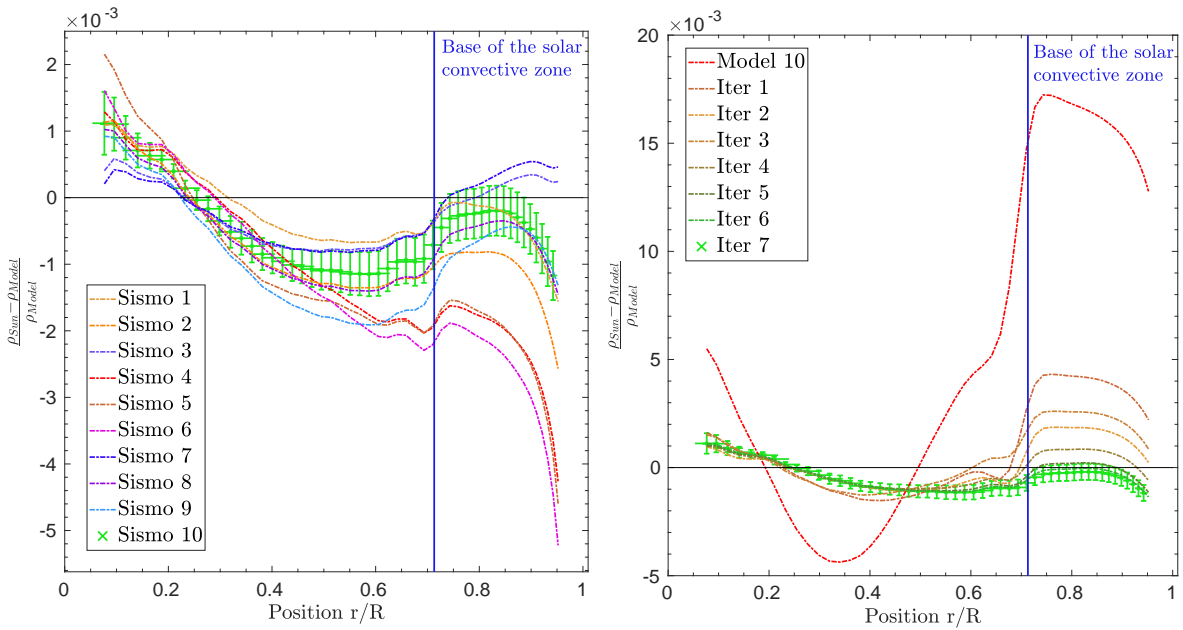}
	\caption{Left panel: Agreement in relative density differences for the reconstructed models using the reference models of Table \ref{tabSTDModels} as initial conditions. Right panel: Illustration of the convergence of the density relative differences for successive iterations of the reconstruction procedure in the case of Model 10.}
		\label{FigRhoAll}
\end{figure*}

The density inversions have been carried out using the ($\rho, \Gamma_{1}$) structural kernels, ensuring that the total solar mass is conserved during the inversion procedure. By using the ($\rho, \Gamma_{1}$) kernels, we ensure an intrinsically low contribution of the cross-term, as the relative variations of $\Gamma_{1}$ are expected to be very small in most of the solar structure. The inversion results for all models are illustrated in the left panel of Fig. \ref{FigRhoAll}. From these results, it appears that most of the models show an excellent agreement in density, of around $0.2\%$ in the radiative layers. The best models show an agreement below $0.1\%$ throughout most of the solar structure and the worst offenders showing discrepancies as high as $0.25\%$ in the deep layers. These are also the models showing the large discrepancies in sound speed discussed earlier and thus, the disagreements we find are actually due to the fact that these models do not fit at best the position of the base of the convective zone of $0.713\pm 0.001~\mathrm{R}_{\odot}$. As discussed in Sect. \ref{SecSoundSpeed}, this means that a second selection can be performed based on the position of the BCZ. This, however, does not have any implication regarding abundances but solely constrains further the behaviour of the Ledoux discriminant around the BCZ and thus has strong implications on the properties of the macroscopic mixing in those layers. However, due to the lack of resolution in the inversion procedure, a definitive answer at the BCZ will likely be the result of non-linear inversions adapted to sample steep gradients of the function to be determined from the seismic data. 

Taking a look at the right panel of Fig. \ref{FigRhoAll}, we can see that the first reconstruction step leads to a significant improvement of the density profile. Unlike the sound-speed inversion, the second and third step lead to smaller improvement of the inversion results in radiative layers. This is also seen in the entropy proxy inversion, where the leap after the first reconstruction step is mainly due to the corrections of the discrepancies around $0.6~\mathrm{R}_{\odot}$ but providing a better overall agreement in the radiative region is mainly due to finer corrections at higher temperatures. This also consistent with the results of \citet{Buldgen2019}, where a localized change of the mean Rosseland opacity could provide some improvement in sound speed and Ledoux discriminant, but was not enough to provide an excellent agreement regarding the entropy proxy inversion. 

\subsection{Frequency-separation ratios}\label{SecRatios}

As a last verification step of the reconstruction procedure, we take a look at the so-called frequency-separation ratios defined in \citet{RoxburghRatios} of the $6$ of our reconstructed solar models. These quantities are defined as the ratio of the so-called small frequency separation over the large frequency separation ratio as follows
\begin{align}
r_{02}=\frac{\nu_{0,n}-\nu_{2,n-1}}{\nu_{1,n}-\nu_{1,n-1}} \\
r_{13}=\frac{\nu_{1,n}-\nu_{3,n-1}}{\nu_{0,n+1}-\nu_{0,n}},
\end{align}
with $\nu_{\ell,n}$ the frequency of radial order $n$ and degree $\ell$. 

\begin{figure*}
	\centering
		\includegraphics[width=0.95\linewidth]{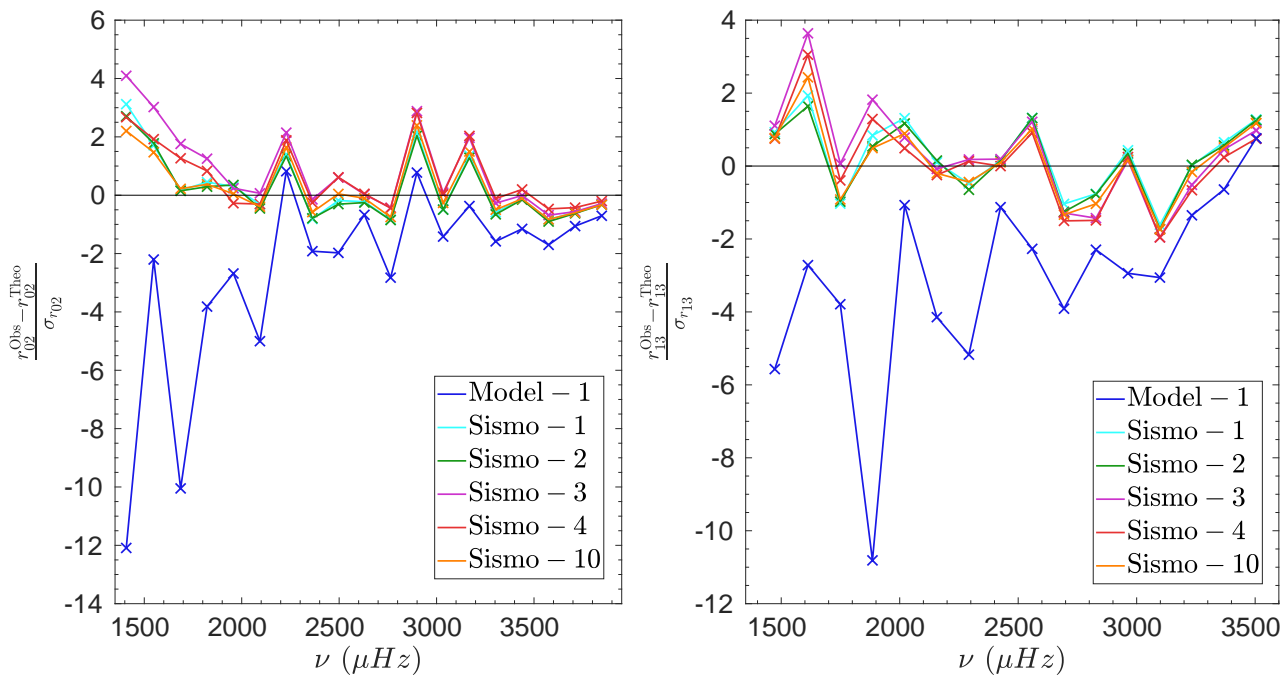}
	\caption{Agreement in frequency-separation ratios of low-degree p-modes for $6$ reconstructed models using the corresponding references of Table \ref{tabSTDModels} as initial conditions as well as reference Model $1$ shown as comparison.}
		\label{FigRatiosSismo}
\end{figure*}

In Fig. \ref{FigRatiosSismo}, we plot the difference between the observed frequency-separation ratios and those of our seismic models, normalized by their $1\sigma$ uncertainties. As can be seen, the agreement for all models is excellent. From the comparison with the results of Model $1$ in blue, we can see that the improvement in the agreement is quite significant. This is no surprise, as the reconstruction is based on reproducing the Ledoux discriminant inversions, which is closely linked to the sound-speed gradient and thus to the frequency-separation ratios, following the asymptotic developments of \citet{Shibahashi1979} and \citet{Tassoul1980}. This also means that the frequency-separation ratios, just as any other classic helioseismic investigation such as structural inversions, are by no means direct measurement of the chemical abundances in the deep radiative layers. As a consequence, they cannot be used to advocate the use of one or the other abundance table for the construction of solar models.   

\section{Impact of reference models}\label{SecDisc}

While the reconstruction procedure leads to a very similar Ledoux discriminant profile for a wide range of initial model properties, it might not be fully justified to say that it is completely independent of the initial conditions of the procedure. In the previous sections, we have discussed how the position of the BCZ could affect the final agreement for sound-speed, density and entropy proxy inversions.

However, this is an impact that can be easily measured from the combination of multiple structural inversions and lead to some extent to an additional selection of the optimal seismic structure of the Sun. Other effects, such as the selected dataset or the assumed behaviour of the surface correction may lead to slight differences at the level of agreement we seek with such reconstruction procedure. Using for example an extended MDI dataset such as the one of \citet{Reiter2015, Reiter2020} may provide ways to further test the robustness of our procedure, as well as better probe the agreement of our seismic model of the Sun in the convective envelope. Indeed, from the comparison of models Sismo 1, Sismo 8 and Sismo 9, we can see that some small variations in the upper envelope properties remain after the reconstruction procedure. The changes in the radiative region between these models are unsurprisingly much more smaller. This implies that additional insight can be gained from using an extended dataset probing in a more stringent manner the upper convective layers, as in \citet{DiMauro2002, Vorontsov13}.

\begin{figure*}
	\centering
		\includegraphics[width=17cm]{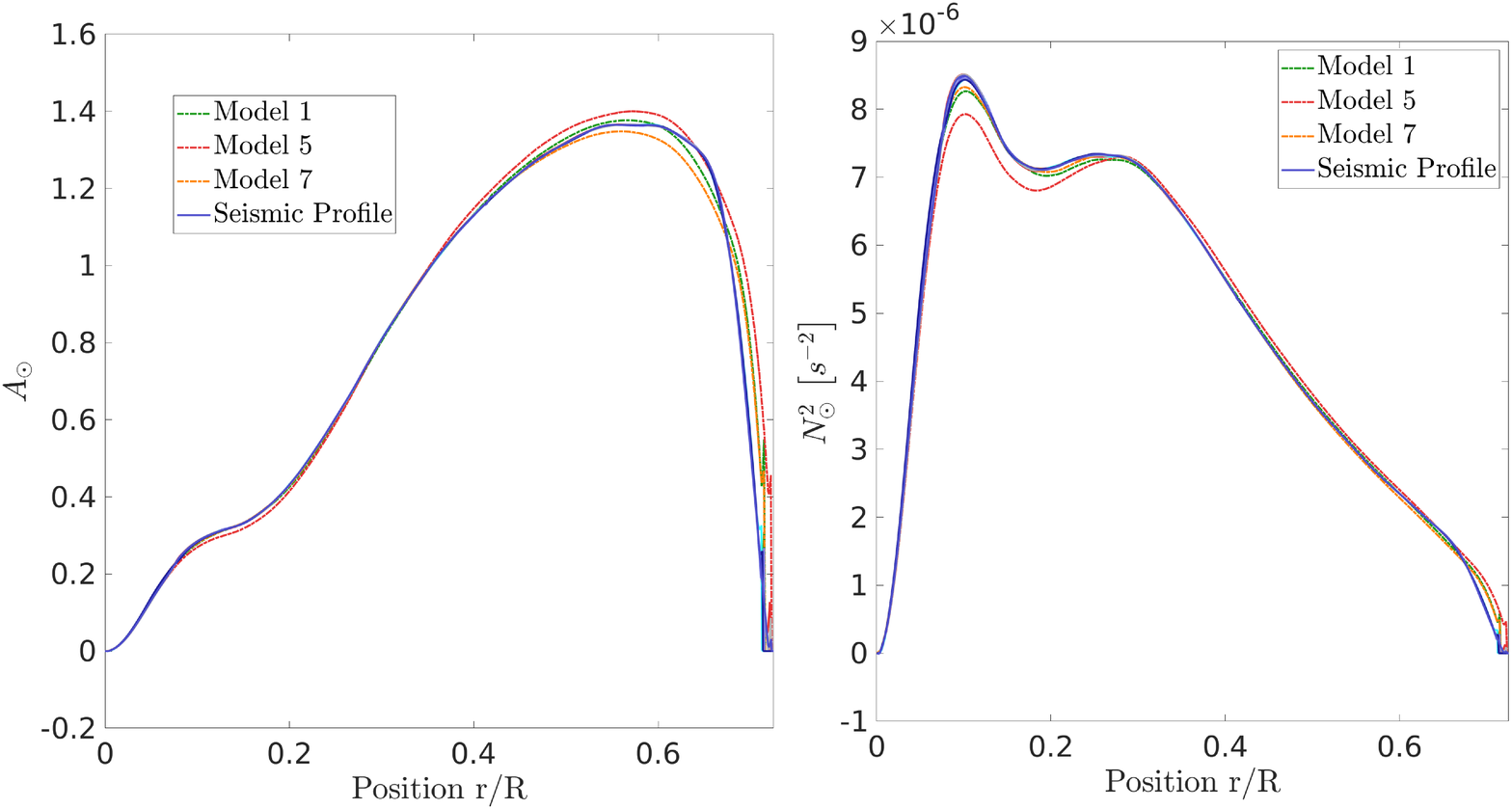}
	\caption{Left panel: Ledoux discriminant profiles from the reconstruction procedure. The dashed lines illustrate the reference profile of some of the models in Table \ref{tabSTDModels} while the continuous lines show the profile on which the procedure converges (using all reference models). Right panel: Same as the left panel but for the Brunt-Väisälä frequency.}
		\label{FigASunRec}
\end{figure*}

\begin{figure}
	\centering
		\includegraphics[width=8cm]{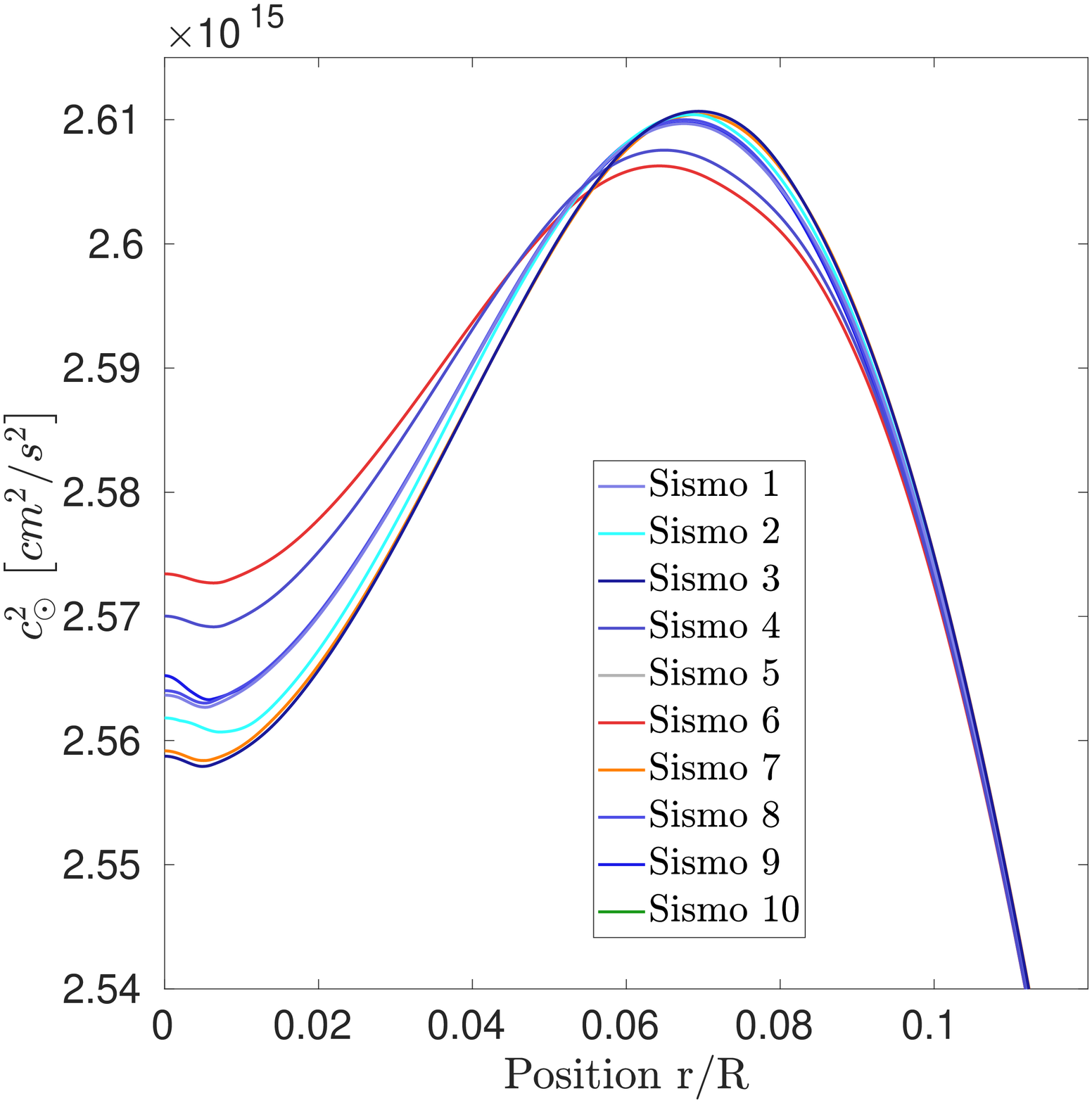}
	\caption{Sound-speed profile in the core region of the reconstructed model, showing the impact of initial conditions in the deep solar core, that is unconstrained by p modes.}
		\label{FigCCoreMod}
\end{figure}

The approach used to connect the regions on which the $A$ corrections are applied to the central regions may also locally impact the procedure. Obviously, the fact that no corrections are applied below $0.08~\mathrm{R}_{\odot}$ leaves a direct mark on the final reconstructed structure. This is illustrated in Fig. \ref{FigASunRec} and Fig \ref{FigCCoreMod} where we plot the Ledoux discriminant, Brunt-Väisälä frequency and sound-speed profiles of all our reconstructed models in the deep solar core. From the inspection of the right panel of Fig. \ref{FigASunRec}, we can understand better the behaviour of the period spacing changes of table \ref{tabSpacing}. Indeed, Model 1 and 7 show some minor changes in asymptotic period spacing, while the spacing of Model 5 is significantly corrected. This is simply due to the fact that Model 1 and 7 reproduce much better the Brunt-Väisälä frequency of the Sun in the deep radiative layers. However, as we lack constraints below $0.08~\mathrm{R}_{\odot}$, we cannot state with full confidence that the observed period spacing will lay within the $25~s$ range we find. 

The optimal approach to lift the degeneracy in the inner core would be to have at our disposal an observed value of the period spacing of the solar gravity modes. Including constraints from neutrino fluxes may already provide additional constraints. However, their main limitation is that in this case, we would have to assume a given composition profile for our solar models, a given equation of state as well as nuclear reaction rates. Taking these constraints into account implies that we use all our current knowledge on the present day solar structure. However, this would be at the expense of adding more uncertainties and "model-dependencies" in the procedure. In their paper, \citet{Shibahashi1996} only made an assumption on $Z(r)$, the distribution of heavy element in the solar interior, but they solved the equation of radiative transfer and thus assumed the mean Rosseland opacity to be known. Given the current uncertainties on radiative opacities, it might be safer to avoid such an assumption, especially at lower temperatures. 

Unravelling the core properties of the Sun would be crucial for the theory of stellar physics. First, it would allow to constrain the angular momentum transport processes acting in solar and stellar radiative zones, as discussed in \citet{Eggenberger2019}. Second, it would demonstrate whether the solar core has undergone intermittent mixing, for example due to out-of-equilibrium burning of $^{3}\mathrm{He}$ \citep{Dilke1972, Unno1975, Gabriel1976} or a prolongated lifetime of its transitory convective core in the early main-sequence evolution. Third, it would also allow us to constrain nuclear reactions rates and their screening factors, as these remain key ingredients of stellar models that are quite uncertain from a theoretical point of view \citep[see e.g.][for a discussion]{Mussack2011A,Mussack2011B}. 

\section{Limitations and uncertainties}\label{SecLimitations}

From the previous sections, we have demonstrated how the inversion of the Ledoux discriminant can be used to build a seismic Sun from successive correction, integration and inversion steps. While the process is quite straightforward and allows us to test the consistency between different inversions of the solar structure it also suffers from intrinsic limitations. 

The first limitation is due to the incomplete information given by the inversion technique. Indeed, since p-modes do not allow us to test solar models below $0.08~R_{\odot}$, we are not able to get a full view of the seismic Sun. While it only represents $8\%$ of the solar structure in radial extent, it also represents $\approx 3\%$ in mass of the solar structure, which actually corresponds to the total mass of the solar convective envelope. This implies that an accurate depiction of the mass distribution inside the Sun can only be hoped for if we observe solar g modes. This is also illustrated from the results of Table \ref{tabSpacing} which shows that the reconstruction only slightly alters the values of the period spacing. Obtaining a precise measurement of that quantity would indeed provide a strong additional selection on the set of seismic models and allow for joint analyses using both neutrino fluxes measurement and helioseismology. 

The second main limitation also stems from the inversion procedure and is linked to the impact of the surface effects. Indeed, as can be seen from sound-speed, entropy proxy and density inversions, the upper layer of the model have not been extensively probed by the inversion procedure. As we only apply corrections in $A$ in the internal radiative layers, this does not render the reconstruction procedure useless. However, this means that the conclusions we can draw from the other inversion procedures used as sanity checks are mostly limited to the lower parts of the envelope and the radiative region. This implies that to better constrain the properties of the solar convective envelope (composition, equation of state, ...), we will have to extend the dataset to higher degree modes. Of course, this means that we will have to be more attentive to the surface effects that can lead to biases in the determined corrections from these higher frequency modes \citep[see e.g.][]{Gough1995}. 

Not applying the corrections in the convective envelope is also a clear limiting factor of our method. While the improvement is significant when comparing the initial models in terms of density and entropy proxy profile, it is also clear that the sound speed profile in the convective envelope is not extensively corrected by our method. This leads our seismic models to be somewhat less performant in terms of sound speed in the convective envelope, while being more performant in the radiative region, especially regarding density and entropy proxy inversions. Consequently, a promising way of obtaining a very accurate picture of the internal structure of the Sun from helioseismology would be to combine our reconstruction technique with those focusing on obtaining optimal models of the solar convective envelope \citep[as shown for example in][]{VorontsovSolarEnv2014}.

In addition to these limitations that are intrinsic to the use of seismic inversions from solar acoustic oscillations, we have also made the hypothesis to keep the $\Gamma_{1}$ profile and the position of the BCZ unchanged in the reconstruction procedure. The BCZ position is not a strong limitation, as it can easily be avoided by ensuring that the reference model agrees with the helioseismically determined value. As we have seen from Fig. \ref{FigASunRec}, this has no impact on the final Ledoux discriminant profile determined by the reconstruction procedure. 

The hypothesis of unchanged $\Gamma_{1}$ can be more severe if we start looking at changes in the thermodynamical quantities of small amplitude. Indeed, \citet{VorontsovSolarEnv2014} have reported measurements of the $\frac{\delta \Gamma_{1}}{\Gamma_{1}}$ profile of a precision of up to $10^{-4}$ in the deepest half part of the convection zone, which is just one order of magnitude bigger than the estimated precision of equation of state interpolation routines \citep{Baturin2019}. This implies that at the magnitude of the relative differences we see in the solar convective envelope, keeping $\Gamma_{1}$ constant might not be a good strategy, but also that we have to become attentive to the intrinsic numerical limitations of the tabulated ingredients of our solar models.

Moreover, intrinsic changes between different equations of state will necessarily lead to slight differences in the determined profiles. Since the $\Gamma_{1}$ profile in the convective envelope is strongly tied to the assumed chemical composition of the envelope, this also means that the properties of the envelope of our seismic models may not be fully consistent. Indeed, the density, pressure and $\Gamma_{1}$ profiles are determined (or fixed) in our approach. Providing consistent profiles of $\mathit{X}$, $\mathit{Z}$ and $\mathit{T}$ for the reconstructed structure might very likely require to assume changes in $\Gamma_{1}$ in the envelope of our models. This is beyond the scope of this study, but will be addressed in the future using the most recent versions of equations of state available \citep{Baturin} using $\Gamma_{1}$ inversions in the solar envelope.

In the radiative layers, the hypothesis of unchanged $\Gamma_{1}$ is less constraining, as one can safely assume that the $A$ corrections will be largely dominated by the contributions from the pressure and density gradients. However, there is still a degeneracy between temperature and chemical composition, that can both lead to changes in pressure and density profiles, even if an equation of state is assumed. This is also a clear limitation of our procedure, but is intrinsic to its seismic nature.

A third limitation of our reconstruction is also linked to the seismic data used. We carried out additional tests using MDI data from \citet{Larson2015} and from the latest release following the fitting methodology of \citet{Korzennik2005}, \citet{Korzennik2008a} and \citet{Korzennik2008b} as well as GOLF data instead of BiSON data for the low degree modes and found that variations of up to $\approx 3\times 10^{-3}$ could be found in the sound speed profiles of the seismic models, while using the latest MDI datasets led to changes of about $\approx 1\times 10^{-3}$ and using GOLF data from \citep{Salabert2015} instead of BiSON data led to similar deviations. The largest differences were however found below $0.08R_{\odot}$, in a region uncontrolled by the reconstruction procedure. This implies that the actual accuracy of the reconstruction is tightly linked to the dataset used.

\section{Conclusion}\label{SecConc}

We have presented a new approach to compute a seismic Sun, taking advantage of the Ledoux discriminant inversions to limit the amount of numerical differentations when computing the solar structure. Our approach allows us to provide a full profile of the Ledoux discriminant for our seismic models. To verify the consistency and robustness of our method, we have checked that it also led to an improvement of other classical helioseismic indicators such as frequency-separation ratios as well as other structural inversions. By selecting models with the exact position of the discontinuity in $A$ determined by helioseismology, the reconstructed models agree with the Sun well within $0.1\%$ for all other structural inversions in the radiative interior. Slightly larger differences can be seen depending on the dataset used to carry out the reconstruction.

Our procedure converges consistently on a unique estimate of the solar Ledoux discriminant within the range of radii on which the inversions are considered reliable. This approach opens new ways of analysing the current uncertainties on the solar temperature gradient, as the main advantage of the Ledoux discriminant is that it is largely dominated by the contribution of the difference between the temperature gradient and the adiabatic temperature gradient in most of the radiative zone. This opens up the possibility to estimate directly the expected opacity modification for a given equation of state and a given chemical composition, providing key insights to the opacity community \citep[see also][]{Gough2004}. This approach provides a complementary way to estimate the required modifications expected to be a key element in solving the current stalemate regarding the solar abundances following their revision by \citet{AGSS09}.

In the regions where the mean molecular weight term of $A$ has a non-negligible contribution (close to the BCZ and in regions affected by nuclear reactions), further degeneracies can be expected and thus larger uncertainties. However, by analysing these effects, we can expect to gain insights on the properties of microscopic diffusion and mixing at the BCZ. Gaining more insights on the behaviour of the discontinuity in the $A$ profile in the tachocline region will require using one of our seismic models as a reference for non-linear RLS inversions as in \citet{corbard99}, allowing for solutions of the inversion displaying larger gradients. This will be done in future studies. 

In the convective envelope, the reconstruction procedure allows to set a good basis for precise determinations of the composition in this region, especially for $\mathit{Z}$, improving on \citet{BuldgenZ}, and tests of the equation of state used in solar and stellar models. Indeed, this requires to avoid as much as possible contaminations by the unavoidable cross-term contributions while still being able to test different equations of states. By using higher $\ell$ modes \citep{Reiter2015, Reiter2020}, we can expect to further test the robustness of our approach in the convective envelope and provide estimates of the physical properties of the Sun in this region. 

Gaining more information on the solar core will very likely be more difficult, as the procedure is intrinsically limited by the solar p-modes. However, combining it with neutrino fluxes may provide a way to further constrain the solar structure. In addition, it may also be useful to predict the expected range of the asymptotic value of the period spacing of g-modes, helping with their detection. This is particularly timely, given the recent discussions in the literature about the potential detection of these modes \citep{Fossat, Fossat2018, Schunker2018, Appourchaux2019, Scherrer2019} and the renewed interest it inspired for the quest to find them. Should this additional seismic constraint become available, it could be easily included in the procedure and open a new era of solar physics. As such, the Sun still remains an excellent laboratory of fundamental physics and the current study provides a new, original way to exploit the information available from the observation of acoustic oscillations.

\section*{Acknowledgements}

We thank the referee for the useful comments that have substantially helped to improve the manuscript. G.B. acknowledges fundings from the SNF AMBIZIONE grant No 185805 (Seismic inversions and modelling of transport processes in stars). P.E. and S. J. A. J. S. have received funding from the European Research Council (ERC) under the European Union's Horizon 2020 research and innovation programme (grant agreement No 833925, project STAREX). This article used an adapted version of InversionKit, a software developed within the HELAS and SPACEINN networks, funded by the European Commissions's Sixth and Seventh Framework Programmes. Funding for the Stellar Astrophysics Centre is provided by The Danish National Research Foundation (Grant DNRF106). We acknowledge support by the ISSI team ``Probing the core of the Sun and the stars'' (ID 423) led by Thierry Appourchaux.

\bibliography{biblioarticleSismo}

\appendix
\section{Numerical details of the reconstruction technique}\label{SecNumDetails}

As mentioned in Sect. \ref{SecMethod}, there are a few technical details related to the reconstruction procedure. Here, we briefly describe some of the numerical aspects used for this study\footnote{The interested reader may find additional descriptions of similar numerical procedures in \citet{ScuflaireOsc}}. 

The reference models are evolutionary models computed using CLES, they contain typically between $1600$ and $2500$ layers, depending on the specifities asked when running the evolutionary sequence. A typical evolutionary sequences for our reference models counts between $250$ and $350$ timesteps. The calibration on the solar parameters is carried out using a Levenberg-Marquardt algorithm, fitting the solar radius, the solar luminosity and the current surface chemical composition at a level of $10^{-5}$ in relative error. 

The reconstruction itself starts with a local cubic spline interpolation in $r^{2}$ using a Hermite polynomial defined by the function value and its derivative at each mesh interval. The derivatives are computed at each point from the analytical formulas of the second order polynomial associated with the interpolation. This interpolation of the reference model is carried out onto a finer grid of typically $4000$ to $5000$ layers (although $3000$ points might be sufficient as long as the sampling is good enough in the central regions if one wishes to compute g modes). This step is made to ensure that the reintegration is done on a fine enough mesh after the corrections. The new grid points are added based on the variations of $r$, $m^{1/3}$, $\log P$, $\log \rho$ and $\Gamma_{1}$.

\begin{figure}
	\centering
		\includegraphics[width=8cm]{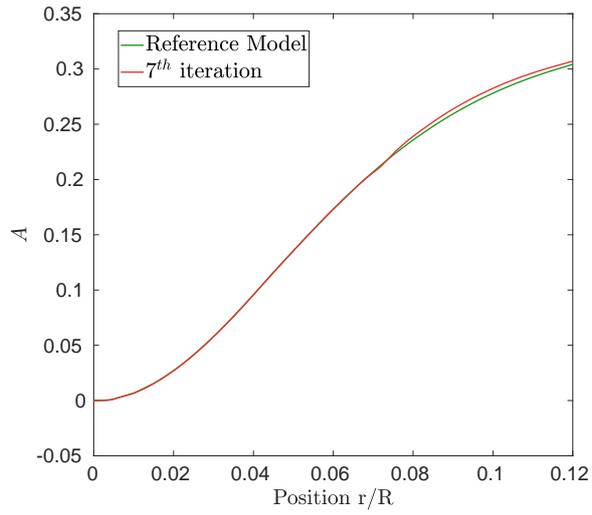}
	\caption{Ledoux discriminant profile of a reconstructed and reference model, showing the lower reconnexion point around $0.08~R_{\odot}$.}
		\label{FigACoreMod}
\end{figure}

The $A$ profile resulting from the inversion is then interpolated on this grid between $0.08~\mathrm{R_{\odot}}$ and the BCZ of the model. No corrections are applied above the BCZ of the model. Below $0.08~\mathrm{R}_{\odot}$, the fact that we do not add the $A$ corrections will lead to an unphysical discontinuity. To avoid this, we reconnect the corrected and the uncorrected profiles by interpolating them on $\approx 20$ layers. This means that the $A$ profile below $0.08~\mathrm{R_{\odot}}$ will remain that of the reference model. An example of such a reconnection is shown in Fig. \ref{FigACoreMod} As described in Sect. \ref{SecMethod}, the $\Gamma_{1}$ profile is kept unchanged throughout the procedure. 

Once the $A$ profile in the radiative zone has been constructed, the model needs to be reintegrated satisfying hydrostatic equilibrium, mass conservation and the boundary conditions in mass and radius. From a formal point of view, the equations to reintegrate the structure are expressed as follows:
\begin{align}
r\frac{d\left(m/r^{3}\right)}{dr}=& 4\pi \rho - 3 \left( m/r^{3}\right), \\
\frac{1}{r}\frac{dP}{dr}=& -G\rho \left(m/r^{3} \right), \\
\frac{d\ln \rho}{dr}=& \left(A/r\right)-\frac{G\rho r}{\Gamma_{1}P}\left(m/r^{3}\right),
\end{align}
with the conditions that $\left(m/r^{3}\right)=\frac{4\pi \rho}{3}$ at $r=0$ and $\left(m/r^{3}\right)=M_{\odot}/R^{3}_{\odot}$ at $r=R_{\odot}$ and $P=P_{\rm{ref}}$ at $r=R_{\odot}$, where $P_{\rm{ref}}$ is the surface pressure of the reference model.

The thermal structure of the model is not taken into account and the equations are discretized on a $4^{th}$ order finite difference scheme. The system is solved with the help of a Newton-Raphson minimization, using the reference model as initial conditions. The final result is a full ``acoustic'' structure: $A$, $\rho$, $P$, $m$, $\Gamma_{1}$ built from the $A$ inversion of a given model. From this ``acoustic'' structure, linear adiabatic oscillation can be computed and the process reiterated until convergence is reached.  

\end{document}